\documentclass[aps,twocolumn,tightline,pra]{revtex4}
\usepackage[colorlinks,citecolor=blue,linkcolor=blue,hyperindex]{hyperref}
\usepackage{amsmath,mathrsfs,amsbsy,color,graphicx,bm,amsthm,amsfonts}
\usepackage{units,bbm,times,dcolumn,amssymb,epsfig,float}
\usepackage{color}
\begin{document}
\title{Quantum speed limit for the maximum coherent state under squeezed environment}
\author{Kang-ying Du$^{1}$, Ya-jie Ma$^{1}$, Shao-xiong Wu$^{1}$\footnote{ sxwu@nuc.edu.cn}, Chang-shui Yu$^{2}$\footnote{ycs@dlut.edu.cn}}

\affiliation{$^1$ School of Science, North University of China, Taiyuan 030051, China\\
$^2$ School of Physics, Dalian University of Technology, Dalian 116024, China}
\date{\today }

\begin{abstract}
The quantum speed limit time for quantum system under squeezed environment is studied. We consider two typical models, the damped Jaynes-Cummings model and the dephasing model. For the damped Jaynes-Cummings model under squeezed environment, we find that the quantum speed limit time becomes larger with the squeezed parameter $r$ increasing and indicates symmetry about the phase parameter value $\theta=\pi$. Meanwhile, the quantum speed limit time can also be influenced by the coupling strength between the system and environment. However, the quantum speed limit time for the dephasing model  is determined by the dephasing rate and the boundary of acceleration region that interacting with vacuum reservoir can be broken when the squeezed environment parameters are appropriately chosen.
\end{abstract}

\maketitle
\section{Introduction}
In the quantum information processing, the evolution of quantum system is of great significance. One may ask what is the minimum time for a quantum state evolute to another state? Mandelstam and Tamm investigated a pure state evolve to its orthogonal state, and found that the shortest time is related to the variance of energy, i.e., $\frac{\pi\hbar}{2\Delta E}$, which can be considered as a extension form of Heisenberg time-energy relation, and is called the MT bound \cite{Mandelstam45}. The shortest time is known as the quantum speed limit time. Under the energy representation, Margolus and Levitin reinvestigated the transition probability amplitude question between two orthogonal quantum pure states, and found that the minimum time is related to the mean value of energy, i.e., $\frac{\pi\hbar}{2\langle E\rangle}$, which is called ML bound \cite{Margolus98}. In order to describe the evolution of quantum state more efficiently, the unified form of quantum speed limit time is given by $\tau_{\text{qsl}}=\max\{\frac{\pi\hbar}{2\Delta E},\frac{\pi\hbar}{2\langle E\rangle}\}$. The evolution of closed system is investigated in diverse ways from the perspective of quantum speed limit \cite{Anandan90,Fleming73,Bhattacharyya83,Vaidman92,Campaioli18, Giovannetti03,Yung06,Jones10,Giovannetti12,Hegerfeldt13}.

With the booming development of quantum information field, the theory of open quantum systems is employed to deal with the dynamical evolution of quantum systems \cite{Breuer07,Breuer16}. Naturely, the concept of quantum speed limit is extended to open quantum systems, and investigated in different ways, such as the quantum Fisher information \cite{Taddei13}, or the relative purity \cite{Campo13}. In Ref. \cite{Deffner13}, a unified form of quantum speed limit time for open quantum systems is proposed by utilizing the von Neumann trace inequality and Cauchy-Schwarz inequality for operators. Inspired by these results, the properties of quantum speed limit for open systems have been reported for both initial pure and mixed states under various distances \cite{Zhang14,Xu14,Wu15,Liu15,Sun15,Zhang15,Liu16,Wei16,Song16, Xu19,Wu20sr,Lu21,Cai17,Zhanglin18,Wu18}. The other aspects of quantum speed limit time were also considered, such as the relationship with the information theory \cite{Pires16,Marvian16} or the quantum control \cite{Campbell17,Xu18,Brody19,Bukov19,Fogarty20, TNXu20,Suzuki20}, the non-Hermite quantum system \cite{Sun19}, the experimental demonstration \cite{Cimmarusti15}. One can see the comprehensive reviews \cite{Frey16,Deffner17jpa} to get more information about the quantum speed limit for open systems. The speed limit in phase space has also been reported \cite{Deffner17,Shiraishi18,Okuyama18,Shanahan18,Nicholson20,Wu20,Hu20}.

The squeezed environment is an important physical resource for quantum information processing \cite{Slusher85,Wu86,Scully97}, it can be used to enhance the parameter estimation, such as the precision of gravitational wave detection \cite{Caves81,Vahlbruch10}. If the environment interacting with the quantum system is  squeezed reservoir, how does the squeezed parameters affect the quantum speed limit for open systems? In this paper, we will investigate the properties of quantum speed limit with squeezed reservoir. Two pedagogical models, the damped Jaynes-Cummings model and the dephasing model, are investigated. Without loss of generality, the initial state is assumed as maximal coherent state. For the damped Jaynes-Cummings model under squeezed environment, the quantum speed limit time indicates symmetrical distribution about the phase parameters $\theta$, and the quantum speed limit bound will be sharper along with the increase of squeezed parameter $r$. In addition, the quantum speed limit time can also be affected by the coupling strength between the quantum system and surrounding environment. In the dephasing model under squeezed environment, the quantum speed limit time is only determined by the dephasing rate, and the acceleration boundary can be broken compared to the case that interacting with vacuum reservoir for appropriate squeezed parameters.

\section{The unifed form of quantum speed limit for open quantum systems}\label{sec2}
In order to measure how close two quantum states are during the evolution, the distance length between the initial state $\rho_0$ and final state $\rho_{\tau}$ is chosen as the Bures angle $\mathcal{L}(\rho _{0},\rho _{\tau })=\arccos [\sqrt{F(\rho _{0},\rho_{\tau })}]$ with fidelity $F(\rho _{0},\rho _{\tau})=(\text{tr}[\sqrt{\sqrt{\rho _{0}}\rho _{\tau }\sqrt{\rho _{0}}}])^{2}$. Through this geometric distance, a unified form of quantum speed limit time for open systems with time-dependent non-unitary dynamical operator $L_{t}(\rho _{t})$ can be obtained following Ref. \cite{Deffner13}.

Employing the von Neumann trace inequality for operators, the ML-type quantum speed limit time for open quantum systems is given as $\tau \geq \max \{ \frac{1}{\Lambda _{\tau }^{\text{op}}},\frac{1}{\Lambda _{\tau }^{\text{tr}}}\} \sin ^{2}\mathcal{L}(\rho ,\rho_{\tau })$ with the quantum evolution rate $\Lambda _{\tau }^{\text{op,tr}}=(1/\tau )\int_{0}^{\tau }dt\Vert L_{t}(\rho _{t})\Vert _{\text{op,tr}}$. Utilizing the Cauchy-Schwarz inequality for operators, the MT-type quantum speed limit time for open quantum systems is $\tau \geq \frac{1}{\Lambda _{\tau }^{\text{hs}}}\sin ^{2}\mathcal{L}(\rho,\rho _{\tau })$ with the quantum evolution rate  $\Lambda _{\tau }^{\text{hs}}=(1/\tau )\int_{0}^{\tau} dt\Vert L_{t}(\rho_{t})\Vert _{\text{hs}}$.  $\Vert \cdot \Vert _{\text{op},\text{tr},\text{hs}}$ means the operator norm, trace norm and Hilbert-Schmidt norm of matrix, respectively. Combining the MT-type and ML-type bounds, a unified form of quantum speed limit time is given as follows
\begin{align}
\tau _{\text{qsl}}=\max \left\{ \frac{1}{\Lambda _{\tau }^{\text{op}}},\frac{1}{\Lambda _{\tau }^{\text{tr}}},\frac{1}{\Lambda _{\tau }^{\text{hs}}}\right\} \sin ^{2}\mathcal{L}(\rho _{0},\rho _{\tau }).\label{eq:inqslty}
\end{align}

Due to the norms of matrix satisfy the inequality $\Vert \cdot\Vert _{\text{op}}\leq \Vert \cdot\Vert _{\text{hs}}\leq \Vert \cdot\Vert _{\text{tr}}$, the ML-type bound of quantum speed limit based on operator norm is tight for open quantum systems. In the following, we will apply the formula (\ref{eq:inqslty}) to the squeezed environment for the damped Jaynes-Cummings model and dephasing model, and investigate the effect of squeezed environment parameters on the properties of quantum speed limit.

\section{The quantum speed limit for damped Jaynes-Cummings model}\label{sec3}
In this section, we will consider the quantum speed limit time for damped Jaynes-Cummings model under squeezed environment. The total Hamiltonian of quantum system and squeezed environment is $H=H_{\text{s}}+H_{\text{env}}+H_{\text{int}}$
with $H_{\text{s}}=\frac{1}{2}\omega_0\sigma_z,H_{\text{env}}=\sum_k\omega_k b_k^{\dag}b_k, H_{\text{int}}=\sum_k(g_k\sigma_+b_k+g_k^*\sigma_-b_k^{\dag})$. The environment is assumed as squeezed vacuum reservoir $\rho_{\text{env}}=\prod_kS_k(r,\theta)\vert0\rangle\langle0\vert S_k^{\dag}(r,\theta)$ with the unitary squeeze operator $S_k(r,\theta)=\exp(\frac{1}{2}re^{-i\theta}b_k^2-\frac{1}{2}re^{i\theta}b_k^{\dag 2})$.

Following Refs. \cite{Wu15pla,Ishizaki08,Wang09}, the non-perturbative master equation through path integral method is given by $\frac{\partial\rho_\text{s}}{\partial t}=-iL_\text{s}\rho_\text{s}-\int_0^td\tau\langle L_\text{int}e^{iL_0(\tau-t)}L_\text{int}e^{-iL_0(\tau-t)}\rangle_\text{env}\rho_\text{s}$ with the super operators $L_\text{s}\rho=[H_\text{s},\rho],L_0\rho=[H_\text{s}+H_{\text{env}},\rho],L_\text{int}\rho=[H_\text{int},\rho]$, and has the following form \cite{Wu15pla}
\begin{align}
\frac{\partial{\rho_s}}{\partial t}=&-(N+1)\alpha(t)(\sigma_+\sigma_-\rho_s-\sigma_-\rho_s\sigma_+)\notag\\
&-(N+1)\alpha^*(t)(\rho_s\sigma_+\sigma_--\sigma_-\rho_s\sigma_+)\notag\\
&-N\alpha(t)(\rho_s\sigma_-\sigma_+-\sigma_+\rho_s\sigma_-)\notag\\
&-N\alpha^*(t)(\sigma_-\sigma_+\rho_s-\sigma_+\rho_s\sigma_-)\notag\\
&+2(\alpha^*(t)M\sigma_+\rho_s\sigma_++\alpha(t)M^*\sigma_-\rho_s\sigma_-),\label{eq:master}
\end{align}
where $N=\sinh^2r$, $M=-\cosh r\sinh r e^{i\theta}$ are given in terms of squeezed parameter $r$ and phase parameter $\theta$.

The structure of squeezed environment interacting with the quantum system is assumed as Lorentz form
\begin{align}
J(\omega)=\frac{\gamma_0}{2\pi}\frac{\lambda^2}{(\omega_0-\omega)^2+\lambda^2},
\end{align}
where the spectral width $\lambda$ is related to correlation time, and coupling strength $\gamma_0$ is determined by the relaxation time.

Without loss of generality, the initial state is chosen as the maximal coherent state $\vert\psi_0\rangle=\frac{1}{\sqrt{2}}(\vert0\rangle+\vert1\rangle)$, the evolved quantum state $\rho(t)$ under squeezed environment is
\begin{align}
\rho(t)=\left(
                   \begin{array}{cc}
                     \rho_{11}(t) & \rho_{10}(t) \\
                     \rho_{01}(t) & 1-\rho_{11}(t) \\
                   \end{array}
                 \right),\label{rho}
\end{align}
where the elements of density matrix (\ref{rho}) are given by
\begin{align*}
 \rho_{11}(t)&=\frac{1}{2}\left(1+\frac{e^{-2\vartheta(t) \cosh 2r}-1}{\cosh 2r}\right),\\
 \rho_{10}(t)&=\frac{1}{4}\left(e^{-e^{2r}\vartheta(t)}(1+e^{i\theta})+e^{-e^{-2r}\vartheta(t)}(1-e^{i\theta})\right),\\
 \rho_{01}(t)&=\rho^*_{10}(t), ~ \rho_{00}(t)=1-\rho_{11}(t)
\end{align*}
with the parameters $\vartheta(t)=\int_0^t d\tau \alpha(\tau)=\frac{\gamma_0}{2}(t+\frac{e^{-\lambda t}-1}{\lambda})$, and $\alpha(t)=\int_0^td\tau\int d\omega J(\omega)e^{i(\omega_0-\omega)(t-\tau)}=\frac{\gamma_0}{2}(1-e^{-\lambda t})$.

The ML-type quantum speed limit time based on operator norm is
\begin{align}
\tau_{\text{qsl}}=\frac{\sin^2\mathcal{L}(\rho_0,\rho_\tau)}{\frac{1}{\tau}\int_0^{\tau}dt\|L_t(\rho_t)\|_{\text{op}}},\label{jcqsl}
\end{align}
where, $\sin^2\mathcal{L}(\rho_0,\rho_{\tau})$ in numerator is $\frac{1}{2}(1-\rho_{01}(t)-\rho_{10}(t))$, and $\|L_t(\rho_t)\|_{\text{op}}$ in denominator is given by $\sqrt{\dot{\rho}_{11}(t)^2+|\dot{\rho}_{10}(t)|^2}$ with the derivative of matrix element
\begin{align*}
&\dot{\rho}_{11}(t)=-e^{-2\cosh(2r)\vartheta(t)}\alpha(t),\\
&\dot{\rho}_{10}(t)=\frac{\alpha(t)}{4}(e^{-2r-e^{-2r}\vartheta(t)}(e^{i\theta}-1)-e^{2r-e^{2r}\vartheta(t)}(e^{i\theta}+1)).
\end{align*}

\begin{figure}[t!]
  \centering
  \includegraphics[width=0.9\columnwidth]{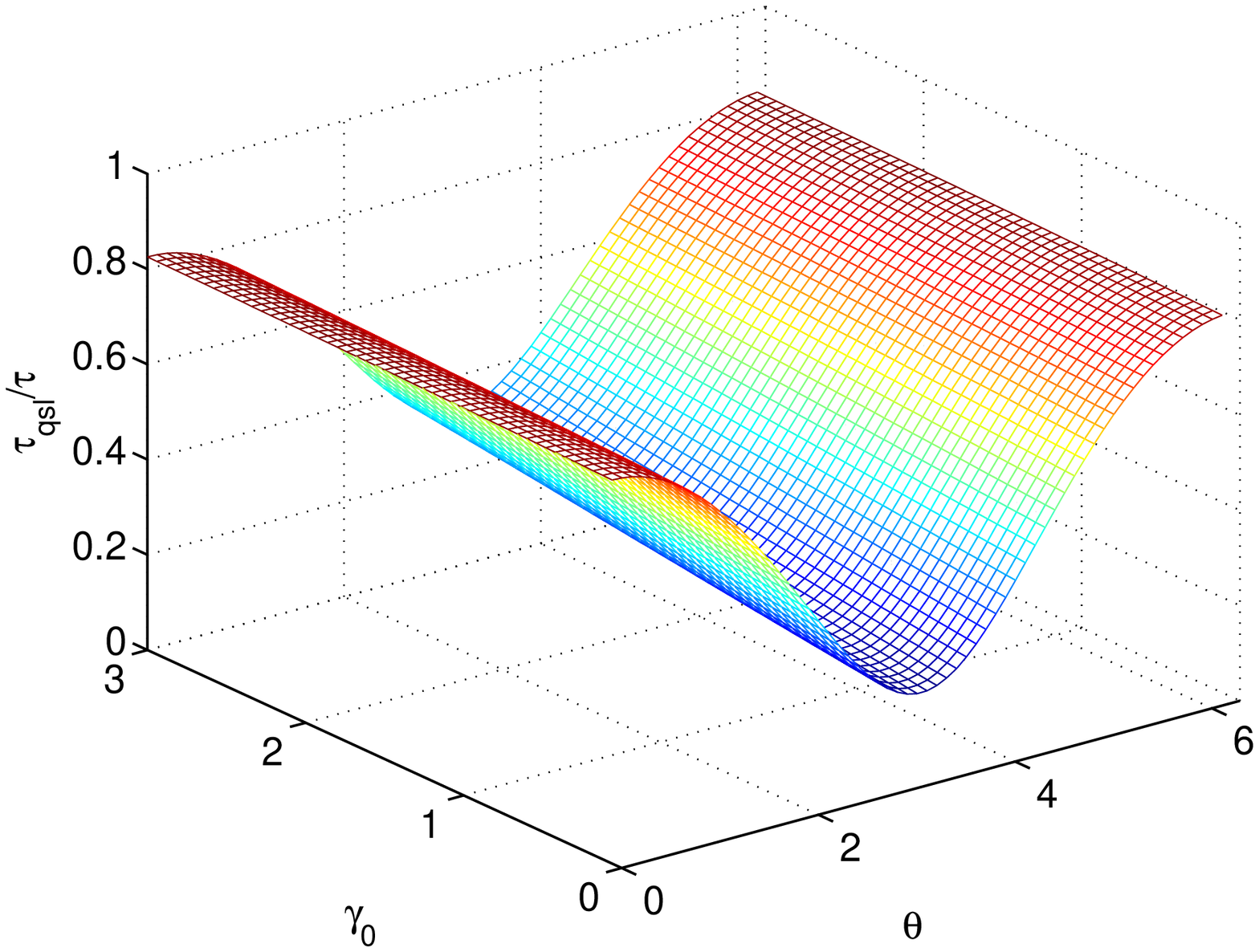}
  \includegraphics[width=0.9\columnwidth]{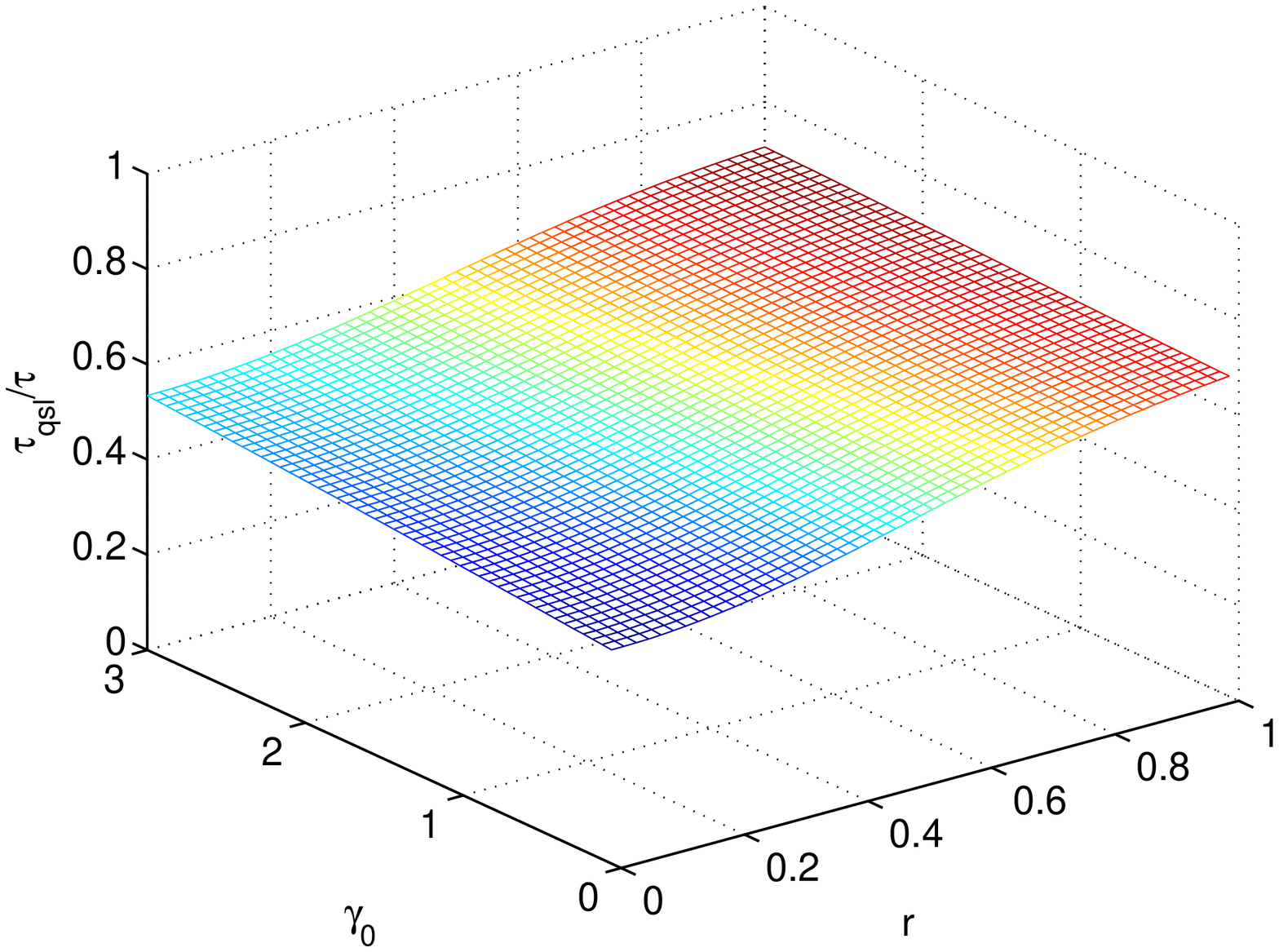}
\caption{(Color online) The function of $\tau_{\text{qsl}}/\tau$ for damped Jaynes-Cummings model under squeezed environment. Panel (a), the ratio $\tau_{\text{qsl}}/\tau$ is varied along with the phase parameter $\theta$ and coupling strength $\gamma_0$. The squeezed parameter is chosen as $r=0.5$. Panel (b), it shows the variation of $\tau_{\text{qsl}}/\tau$ along with the squeezed parameter $r$ and coupling strength $\gamma_0$. The phase parameter is chosen as $\theta=0.5\pi$. In both panels, the spectral width is set as $\lambda=1$ (in units of $\omega_0$) and the actual driving time is chosen as $\tau=1$.}\label{fig1}
\end{figure}

In Fig. \ref{fig1}, the behavior of ratio $\tau_{\text{qsl}}/\tau$ with squeezed environment parameters $r$ and $\theta$ are shown, respectively. In both panels, the environment spectral width is set as $\lambda=1$ (in units of $\omega_0$) and the actual driving time is chosen as $\tau=1$. In Panel (a), it is shown that the variation of quantum speed limit time along with the coupling strength $\gamma_0$ and phase parameter $\theta$. The squeezed parameter is chosen as  $r=0.5$ and the quantum speed limit time shows symmetry about the phase parameter value $\theta=\pi$. The ML-type quantum speed limit time (\ref{jcqsl}) is only comprised of the phase parameter $\theta$ with form $e^{i\theta}$ and its complex conjugate, so it presents periodic characteristics, obviously. When other parameters are fixed, the maximum of quantum speed limit time is located at $\theta=0$. It indicates that the appropriate phases parameter $\theta$ can accelerate the evolution of quantum states under squeezed environment.

Without loss of generality, the phase parameter is chosen as $\theta=0.5\pi$ in Panel (b), and the variation of quantum speed limit time along with the coupling strength $\gamma_0$ and squeezed parameter $r$ are demonstrated. The quantum speed limit time will become larger along with the increasing of squeezed parameter $r$. According to Eq. (\ref{jcqsl}), a possible explanation is that the average evolution velocity of quantum system, i.e., $\Lambda_{\tau}^{\text{op}}$, will become slower when the squeezed parameter is increased, so the bound of quantum speed limit will be tight under squeezed environment.

Even though without Markovian approximation used in non-perturbative master equation (\ref{eq:master}), the variation of environment coupling strength $\gamma_0$ does not reflect the phenomenon of non-Markovianity for open quantum systems \cite{Breuer07,Breuer16}. It maybe caused by different methods to deal with the master equation. However, the quantum speed limit time will be larger along with the coupling strength $\gamma_0$ increasing, which is distinguished with the result in Ref. \cite{Deffner13} even when the squeezing effect is vanished.

\section{The quantum speed limit for dephasing model}\label{sec4}

In this section, we will consider another pedagogical model, the dephasing model, and assume the environment interacting with quantum system is squeezed vacuum state, the total Hamiltonian of quantum system and environment is $H=\frac{1}{2}\omega_0\sigma_z+\sum_k\omega_k b_k^{\dag}b_k+\sum_kg_k\sigma_z(b_k+k^{\dag})$.

The initial quantum state is also assumed as maximal coherent state $|\psi_0\rangle=\frac{1}{\sqrt{2}}(\vert0\rangle+\vert1\rangle)$, and the evolved state is
\begin{align}
\rho(t)=\left(
          \begin{array}{cc}
            \frac{1}{2} & \frac{1}{2}e^{-\gamma(t)} \\
            \frac{1}{2}e^{-\gamma(t)} & \frac{1}{2} \\
          \end{array}
        \right),\label{eqdephasing}
\end{align}
where the dephasing factor $\gamma(t)$ is defined by $\gamma(t)=\int d\omega J(\omega)\frac{1-\cos(\omega t)}{\omega^2}[\cosh(2r)-\cos(\omega t-\theta)\sinh(2r)]$. Without loss of generality, the structure of environment is Ohmic-like spectrum with soft cutoff
\begin{align}
J(\omega)=\eta\frac{\omega^s}{\omega_\text{c}^{s-1}}\exp(-\frac{\omega}{\omega_\text{c}}),
\end{align}
where $\eta$ is the coupling parameter, the cutoff frequency $\omega_\text{c}$ is unity, and $s$ determines the type of environment, i.e., sub-Ohmic environment($s < 1$), Ohmic environment ($s = 1$), and super-Ohmic environment ($s > 1$), respectively. The dephasing factor $\gamma(t)$ in Eq. (\ref{eqdephasing}) can be given analytically \cite{Wu17}
\begin{align}
\gamma(t)=&\frac{\eta}{4}\Gamma(s-1)\{2\cosh(2r)[2-(1+it)^{1-s}-(1-it)^{1-s}]\notag\\
&+e^{-i\theta}\sinh(2r)[1-2(1-it)^{1-s}+(1-2it)^{1-s}]\notag\\
&+e^{i\theta}\sinh(2r)[1-2(1+it)^{1-s}+(1+2it)^{1-s}]\}.
\end{align}

The ML-type quantum speed limit time based on operator norm for dephasing model under squeezed environment can be given as following
\begin{align}
\tau_{\text{qsl}}=\frac{\sin^2\mathcal{L}(\rho_0,\rho_{\tau})}{\frac{1}{\tau}\int_0^{\tau}dt\|L_t(\rho_t)\|_{\text{op}}},\label{qslspin}
\end{align}
where, $\sin^2\mathcal{L}(\rho_0,\rho_{\tau})=\frac{1}{2}(1-e^{-\gamma(t)})$, and $\|L_t(\rho_t)\|_{\text{op}}=\frac{1}{2}|\gamma'(t)e^{-\gamma(t)}|$ with the dephasing rate
\begin{align}
\gamma'(t)=&\frac{i\eta}{2}\Gamma(s)\{\cosh(2r)[(1+it)^{-s}-(1-it)^{-s}]\notag\\
&+e^{-i\theta}\sinh(2r)[(1-2it)^{-s}-(1-it)^{-s}]\notag\\
&+e^{i\theta}\sinh(2r)[(1+it)^{-s}-(1+2it)^{-s}]\}.\label{gammadian}
\end{align}

\begin{figure}[t!]
  \centering
  \includegraphics[width=0.9\columnwidth]{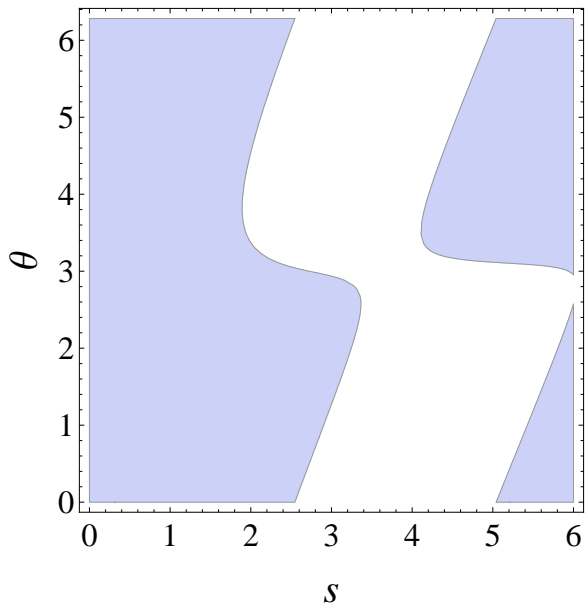}
\caption{(Color online) The sign of dephasing rate $\gamma'(t)$ in Eq. (\ref{gammadian}). The purple region means $\gamma'(t)>0$ and the white region is $\gamma'(t)<0$. The horizontal axis is Ohmic spectrum parameter $s$ and the vertical axis is phase parameter $\theta$. The squeezed parameter $r$ is set as $1$, and the actual driven time is chosen as  $\tau=3$.}\label{fig2}
\end{figure}

The quantum speed limit time for dephasing model under squeezed reservoir (\ref{qslspin}) is only determined by the dephasing rate $\gamma'(t)$, and it shows acceleration features when the dephasing rate $\gamma'(t)$ is negative, which is similar to the result in Ref. \cite{Wu15}. The sign of dephasing rate $\gamma'(t)$ is depicted in Fig. \ref{fig2}, the squeezed parameter is chosen as $r = 1$ and the driven time is chosen as $\tau = 3$. The region of purple means that the value of $\gamma'(t)$ is positive, while the white region means the corresponding negative range. Different to Ref. \cite{Wu15}, the boundary that the dephasing rate $\gamma'(t)=0$ is a curve related to the squeezed parameters $r$ and $\theta$ in Fig. \ref{fig2}. The evolution of quantum system can be accelerated in the white region. Especially, the quantum speed limit bound under vacuum reservoir does not show speedup phenomena when $s\lesssim2.5$ in Ref. \cite{Wu15}. Under the squeezed environment, this restricted condition can be broken when choosing appropriate squeezed environment parameters.

\section{Discussion and conclusion}
We have investigated the quantum speed limit for open quantum system under squeezed environment. Two typical models, i.e., the damped Jaynes-Cummings model and the dephasing model, are considered, and the initial states are chosen as maximal coherent state for both models. For the damped Jaynes-Cummings model, the quantum speed limit time is larger along with the squeezed parameter $r$ increasing and shows symmetry about the phase parameter $\theta$. While, for the dephasing model, the quantum speed limit time is determined by the dephasing rate $\gamma'(t)$ and the acceleration boundary that interacting with vacuum reservoir can be broken under the squeezed reservoir for appropriate squeezed parameters. We expect that our research has contribution to understand the evolution of quantum states in squeezed environments.

\section*{ACKNOWLEDGMENT}
This work was supported by the National Natural Science Foundation of China (Grant No. 11775040), Scientific and Technological Innovation Programs of Higher Education Institutions in Shanxi (Grant No. 2019L0527).

\end{document}